\documentclass[english]{IEEEtran}
\usepackage[T1]{fontenc}
\usepackage[latin9]{inputenc}
\usepackage{amsthm}
\usepackage{amsmath}
\usepackage{amssymb}
\usepackage{stmaryrd}
\usepackage{stackrel}
\usepackage{graphicx}

\makeatletter

\providecommand{\tabularnewline}{\\}

\pdfoutput=1
\theoremstyle{plain}
\ifx\thechapter\undefined
\newtheorem{thm}{\protect\theoremname}
\else
\newtheorem{thm}{\protect\theoremname}[chapter]
\fi
\theoremstyle{plain}
\newtheorem{lem}[thm]{\protect\lemmaname}

\makeatother

\usepackage{babel}
\providecommand{\lemmaname}{Lemma}
\providecommand{\theoremname}{Theorem}

\begin{document}

\title{The Broadcast Channel with Degraded Message Sets and Unreliable Conference }

\author{Dor Itzhak and Yossef Steinberg}

\IEEEspecialpapernotice{\emph{Department of Electrical Engineering}\\
\emph{Technion - Israel Institute of Technology}\\
\emph{Haifa 32000, Israel}\\
\emph{doritz@campus.technion.ac.il ~~~ysteinbe@ee.technion.ac.il}}
\maketitle
\begin{abstract}
As demonstrated in many recent studies, cooperation between users
can greatly improve the performance of communication systems. Most
of the works in the literature present models where all the users
are aware of the resources available for cooperation. However, the
scenario where cooperation links are sometimes unavailable or that
some users cannot be updated whether the cooperation links are present
or not, is more realistic in today's dynamic ad-hoc communication
systems. In such a case we need coding schemes that exploit the cooperation
links if they are present, and can still operate if cooperation is
not possible. In this work we study the general broadcast channel
model with degraded message sets and cooperation links that may be
absent, and derive it's capacity region under such uncertainty conditions.\end{abstract}


\section{Introduction}

Coding schemes that utilize cooperation links between users in a communication
network can greatly improve the communication performance of the network.
Unfortunately, in modern ad-hoc communication systems the availability
of cooperation links is not guaranteed a priori. A typical scenario
in such systems is that the users are aware of the possibility that
some nodes in the network will serve as relays or helpers, but their
help is unreliable and cannot be guaranteed a priori. Therefore, it
is desired to derive coding schemes that exploit cooperation when
it is available, but can still operate when they are not.\let\thefootnote\relax\footnote{This work was supported by the Israel Science Foundation (grant No. 1285/16).}

The broadcast channel is one of the main building blocks of multiuser
communication networks, and as such draws much research efforts. The
physically degraded broadcast channel with conferencing decoders was
introduced and studied in \cite{Dabora2004,Dabora2006}, and a related
model with relay channel in \cite{Liang2007,YingbinLiang}. The non-degraded
BC with degraded message sets and a conference link was presented
and studied in \cite{Steinberg2015}. Regarding the more realistic
scenario, when the conference is unreliable, a physically degraded
BC with a conference that may be absent was suggested and studied
in \cite{Steinberg2014}, and later also in \cite{HuleihelSteinberg2016}.

In this work we extend the results of \cite{Steinberg2014,HuleihelSteinberg2016}
to the general two user BC with degraded message sets and unreliable
conference, building on \cite{Steinberg2015}. Note that, while stochastically
degraded BC is a well accepted model, that can be justified in some
realistic scenarios (e.g, the scalar Gaussian BC), a physically degraded
model is much harder to justify. Hence, with realistic cooperation
problems in mind, the importance of the results presented here lie
mainly in the extension of \cite{Steinberg2014} and \cite{HuleihelSteinberg2016},
to the more general channel model of \cite{Steinberg2015}, getting
rid of the degradedness assumption. In Section II we define the model
and in Section III we present the capacity region. Sketches of the
proofs are given in Section IV.

\section{System Model and Definitions}

\textbf{Definition 1.} An $\left(n,\mu_{0},\mu'_{0},\mu_{1},v_{1},\epsilon\right)$
code for the DM-BC with degraded message sets and unreliable conference
link consists of index sets ${\cal M}_{i}=\left\{ 1,2,...,\mu_{i}\right\} ,i=0,1$,
${\cal M}'_{0}=\left\{ 1,2,...,\mu'_{0}\right\} $ and ${\cal N}_{1}=\left\{ 1,2,...,v_{1}\right\} $,
an encoder mapping:
\[
f:{\cal M}_{0}\vartimes{\cal M}'_{0}\vartimes{\cal M}_{1}\longrightarrow{\cal X}^{n}
\]
a conference mapping:
\[
\phi:{\cal Y}_{1}^{n}\longrightarrow{\cal N}_{1}
\]
and three decoder mappings
\begin{eqnarray*}
g_{1} & : & {\cal Y}_{1}^{n}\longrightarrow{\cal M}_{0}\vartimes{\cal M}'_{0}\vartimes{\cal M}_{1}\\
g_{2} & : & {\cal Y}_{2}^{n}\longrightarrow{\cal M}_{0}\\
g'_{2} & : & {\cal Y}_{2}^{n}\vartimes{\cal N}_{1}\longrightarrow{\cal M}'_{0}
\end{eqnarray*}
such that the average probabilities of error if the conference link
is present or not, denoted by $P_{e}^{'\left(n\right)}$ and $P_{e}^{\left(n\right)}$
respectively, do not exceed $\epsilon$. The common message $M_{0}$,
the residual common message $M'_{0}$ and the private message $M_{1}$,
are uniformly distributed on the index set ${\cal M}_{0}\vartimes{\cal M}'_{0}\vartimes{\cal M}_{1}$.
The probabilities of error for the two cases are given by:
\begin{align*}
P_{e}^{'\left(n\right)} & ={\textstyle \frac{1}{\mu_{0}\mu'_{0}\mu_{1}}}\underset{{\scriptstyle m_{0},m'_{0},m_{1}}}{\sum}P_{Y_{1}Y_{2}\vert X}\left\{ S'{}_{e}\vert f\left(m_{0},m'_{0},m_{1}\right)\right\} \\
P_{e}^{\left(n\right)} & ={\textstyle \frac{1}{\mu_{0}\mu'_{0}\mu_{1}}}\underset{{\scriptstyle m_{0},m'_{0},m_{1}}}{\sum}P_{Y_{1}Y_{2}\vert X}\left\{ S{}_{e}\vert f\left(m_{0},m'_{0},m_{1}\right)\right\} 
\end{align*}
where the sets $S_{e}$,$S'_{e}$ are defined as
\begin{align*}
S{}_{e} & \triangleq\left\{ \left(\boldsymbol{y}_{1},\boldsymbol{y}_{2}\right):g_{1}(\boldsymbol{y}_{1})\neq\left(m_{0},m'_{0},m_{1}\right)\,or\,g_{2}(\boldsymbol{y}_{2})\neq m_{0}\right\} \\
S'{}_{e} & \triangleq S{}_{e}\cup\left\{ \left(\boldsymbol{y}_{1},\boldsymbol{y}_{2}\right):g'_{2}(\boldsymbol{y}_{2},\phi\left(\boldsymbol{y}_{1}\right))\neq m'_{0}\right\} 
\end{align*}
and for notational convenience, the dependence of $S_{e}$ and $S'_{e}$
on the messages is dropped. The conference rate $C_{1}$ and the communication
rates $\left(R_{0},R'_{0},R_{1}\right)$ are defined as:
\[
\begin{array}{cccc}
C_{1}=\dfrac{\log\left(v_{1}\right)}{n} & R_{k}=\dfrac{\log\left(\mu_{k}\right)}{n},k=0,1 & R'_{0}=\dfrac{\log\left(\mu'_{0}\right)}{n}\end{array}
\]
A rate triple $\left(R_{0},R'_{0},R_{1}\right)$ is said to be achievable,
if for any $\epsilon>0$, $\gamma>0$ and sufficiently large $n$,
there exists an $\left(n,2^{n\left(R_{0}-\gamma\right)},2^{n\left(R'_{0}-\gamma\right)},2^{n\left(R_{1}-\gamma\right)},2^{n\left(C_{1}+\gamma\right)},\epsilon\right)$
code for the DM-BC with degraded message sets and unreliable conference
link. The capacity region of the DM-BC with degraded message sets
and unreliable conference link of capacity $C_{1}$ is the closure
of the set of achievable rates $\left(R_{0},R'_{0},R_{1}\right)$
for a given $C_{1}$, and is denoted by ${\cal C}$.

\begin{figure}
\includegraphics[scale=0.6]{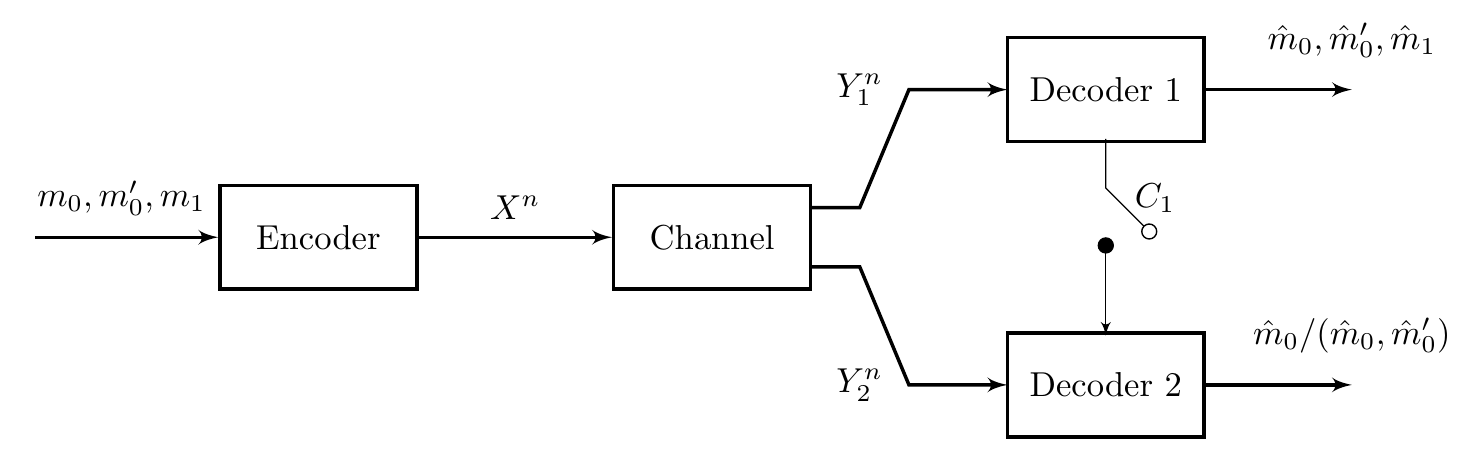}

\protect\caption{The broadcast channel with degraded message sets and unreliable conferencing
decoders}
\end{figure}

\section{Capacity region for the BC with degraded message sets and unreliable
conference}

Let ${\cal R}^{i}$ be the set of all rate triples $\left(R_{0},R'_{0},R_{1}\right)$
satisfying:\begin{subequations} \label{eq:IB}
\begin{align}
R_{0}\le & I\left(U;Y_{2}\right)\\
R'_{0}\le & I\left(V;Y_{2}\vert U\right)+C_{1}\\
R_{1}\le & I\left(X;Y_{1}\vert UV\right)\\
R'_{0}+R_{1}\le & I\left(X;Y_{1}\vert U\right)\label{eq:IB4}\\
R_{0}+R'_{0}+R_{1}\le & I\left(X;Y_{1}\right)\label{eq:IB5}
\end{align}
\end{subequations}for some joint distribution $p\left(u,v,x\right)p\left(y_{1},y_{2}\vert x\right)$.
Consider some special cases of this region:

Case 1. \emph{BC with degraded message sets} - In this model, there
is no residual common message, i.e. $R'_{0}=0$, and no intention
to use the conference link. Here we choose $V=\emptyset$ to get the
capacity region of the BC with degraded message sets first solved
in \cite{Korner1977}. Denote this region by ${\cal C}_{BC}$.

Case 2. \emph{BC with conferencing decoders and degraded message sets}
- In this model, there is no common message, i.e. $R_{0}=0$, so there
is no intention to communicate if the conference link is absent. Here
we choose $U=\emptyset$ to get the capacity region of the non-degraded
BC with degraded message sets and conferencing decoders presented
in \cite{Steinberg2015}. Denote this region by ${\cal C}_{BCC}$
(BCC for BC with Conference).

Case 3. \emph{BC with unreliable conference link and common messages
only} - This special case not yet treated before and presented here
for the sake of completeness. In this model, there is no private message,
i.e. $R_{1}=0$, so there are only common messages to communicate.
Here we choose $V=X$:
\begin{align*}
R_{0}\le & I\left(U;Y_{2}\right)\\
R'_{0}\le & min\left\{ I\left(X;Y_{1}\vert U\right),I\left(X;Y_{2}\vert U\right)+C_{1}\right\} \\
R_{0}+R'_{0}\le & I\left(X;Y_{1}\right)
\end{align*}
for some joint distribution $p\left(u,x\right)p\left(y_{1},y_{2}\vert x\right)$.
Denote this region by ${\cal C}_{noR_{1}}$. We claim that this is
the capacity region for this channel with $R_{1}=0$. For detailed
proof see Part D of Section IV.

Case 4. \emph{Degraded BC with conference link that may be absent}
- Assume that decoder 2 is physically degraded with respect to decoder
1. In that case, the region coincides with the capacity region of
the degraded BC with conference link that may be absent presented
in \cite{Steinberg2014}, by treating the private messages as degraded
message sets as decoder 1 can recover the private message intended
to decoder 2.\\
Let ${\cal R}^{o}$ be the set of all rate triples $\left(R_{0},R'_{0},R_{1}\right)$
satisfying:\begin{subequations} \label{eq:UB}
\begin{align}
R_{0}\le & I\left(U;Y_{2}\right)\label{eq:UB1}\\
R_{0}+R'_{0}\le & I\left(UV;Y_{2}\right)+C_{1}\label{eq:UB2}\\
R_{0}+R'_{0}+R_{1}\le & I\left(X;Y_{1}\right)\label{eq:UB3}\\
R_{0}+R'_{0}+R_{1}\le & I\left(U;Y_{2}\right)+I\left(X;Y_{1}\vert U\right)\label{eq:UB4}\\
R_{0}+R'_{0}+R_{1}\le & I\left(UV;Y_{2}\right)+C_{1}+I\left(X;Y_{1}\vert UV\right)\label{eq:UB5}\\
R_{0},R'_{0},R_{1}\ge & 0\label{eq:UB6}
\end{align}
\end{subequations}for some joint distribution $p\left(u,v,x\right)p\left(y_{1},y_{2}\vert x\right)$.
\begin{thm}
\label{thm:BCCMBAall_ISIT}For the DM-BC with degraded message sets
and unreliable conference link, the capacity region is given by
\[
{\cal C}={\cal R}^{o}={\cal R}^{i}
\]

\end{thm}
\emph{Outline of the proof:} The region ${\cal R}^{i}$ is an achievable
region, i.e ${\cal R}^{i}\subseteq{\cal C}$; the proof of the direct
part is quite similar to the proof in \cite{Steinberg2014} - the
encoder utilizes superposition coding and use binning for the residual
message. A sketch of the converse part is given in Section IV. To
complete the proof we will show the equivalence of those regions,
i.e. ${\cal R}^{i}={\cal R}^{o}$.

\textbf{Example.} Consider the AWGN BC with degraded message sets
and unreliable link with capacity $C_{1}$:
\begin{eqnarray*}
Y_{i}=X+Z_{i} & Z_{i}\thicksim{\cal N}\left(0,N_{i}\right) & i=1,2
\end{eqnarray*}
where $N_{2}>N_{1}$, the noise signals $Z_{1}$ and $Z_{2}$ are
independent and an input power constraint $E\left[X^{2}\right]\le P$.
Denote the classical AWGN capacity by:
\[
{\cal C}\left(x\right)={\textstyle \frac{1}{2}}\log\left(1+x\right)
\]
The capacity region for this model is given by the set of all rate
triples $\left(R_{0},R'_{0},R_{1}\right)$ satisfying:\begin{subequations} \label{eq:GCapacity}
\begin{eqnarray}
R_{0} & \le & {\cal C}\left({\textstyle \frac{\alpha_{0}P}{N_{2}+\left[\alpha'_{0}+\alpha_{1}\right]P}}\right)\label{eq:GCapacity1}\\
R'_{0} & \le & {\cal C}\left({\textstyle \frac{\alpha'_{0}P}{N_{2}+\alpha_{1}P}}\right)+C_{1}\label{eq:GCapacity2}\\
R_{1} & \le & {\cal C}\left({\textstyle \frac{\alpha_{1}P}{N_{1}}}\right)\label{eq:GCapacity3}\\
R'_{0}+R_{1} & \le & {\cal C}\left({\textstyle \frac{\left[\alpha'_{0}+\alpha_{1}\right]P}{N_{1}}}\right)\label{eq:GCapacity4}
\end{eqnarray}
\end{subequations}where $\alpha_{0},\alpha'_{0},\alpha_{1}\ge0$
and $\alpha_{0}+\alpha'_{0}+\alpha_{1}=1$. For detailed proof see
Part E of Section IV.

\section{Proofs}

We will start with the proof of equivalence, i.e. ${\cal R}^{o}={\cal R}^{i}$,
then we give the converse and the direct part. In order to prove\emph{
}${\cal R}^{o}={\cal R}^{i}$ recall some definitions and lemmas from
convex analysis. See \cite{Rockafellar} for the definition of extreme
points, and \cite{Henk} for the properties of polytopes.\textbf{}\\
\textbf{Definitions:} Let $S$ be a set of points in $\mathbb{R}^{k}$,
i.e. $S\subseteq\mathbb{R}^{k}$. 
\begin{itemize}
\item A point $\mathbf{x}\in S$ is called an \emph{extreme point }of $S$
if there do not exist $\mathbf{x}_{1},\mathbf{x}_{2}\in S$ and $\lambda\in\left(0,1\right)$
where $\mathbf{x}_{1}\neq\mathbf{x}_{2}$, such that $\mathbf{x}=\lambda\mathbf{x}_{1}+\left(1-\lambda\right)\mathbf{x}_{2}$.
\item Let $ext\left(S\right)$ be the set of all extreme points of $S$.
\item Let $conv\left(S\right)$ be the convex hull of $S$.
\item Let $\mathbf{A}\in$$\mathbb{R}^{m\times k}$ , $\mathbf{b}\in$$\mathbb{R}^{m}$.
Then $P\triangleq\left\{ \mathbf{x}\in\mathbb{R}^{k}\vert\mathbf{A}\mathbf{x}\le\mathbf{b}\right\} $
is called a convex ${\cal H}$-generalized polytope (${\cal H}$ for
half-space). If in addition $P$ is bounded it called a convex ${\cal H}$-polytope.
\item Let $\left\{ \mathbf{x}_{1},...,\mathbf{x}_{l}\right\} $ be a set
of points in $\mathbb{R}^{k}$. Then $P\triangleq conv\left(\left\{ \mathbf{x}_{1},...,\mathbf{x}_{l}\right\} \right)$
is called a convex ${\cal V}$-polytope (${\cal V}$ for vertex).\end{itemize}
\begin{lem}
\label{lem:conv(ext)}Let $S\subseteq\mathbb{R}^{k}$ be a compact
convex set, then
\[
S=conv\left(ext\left(S\right)\right)
\]

\end{lem}
Lemma \ref{lem:conv(ext)} is common in convex analysis, see Corollary
18.5.1 on page 167 in \cite{Rockafellar}.
\begin{lem}
\label{lem:polytopes representation}\textbf{Main theorem of polytopes
theory} - Let $P\subseteq\mathbb{R}^{k}$ be a convex ${\cal H}$-polytope,
then it has an equivalent representation as a convex ${\cal V}$-polytope
by it's own vertices.
\end{lem}
Lemma \ref{lem:polytopes representation} can be found in \cite{Henk},
and it's proof can be found in \cite{Rockafellar} Theorem 19.1 and
Corollary 19.1.1.

\subsection{Equivalence proof}

It is straight forward to prove that ${\cal R}^{i}\subseteq{\cal R}^{o}$
- for every $p\left(u,v,x\right)$, the inequalities satisfied in
${\cal R}^{i}$ imply that the inequalities in ${\cal R}^{o}$ are
also satisfied. In order to prove ${\cal R}^{o}\subseteq{\cal R}^{i}$
we cannot use the last argument. Instead we will prove that $ext\left({\cal R}^{o}\right)\subseteq{\cal R}^{i}$
and then use Lemma \ref{lem:conv(ext)} and the convexity property
of those regions to conclude:
\[
{\cal R}^{o}=conv\left(ext\left({\cal R}^{o}\right)\right)\subseteq conv\left({\cal R}^{i}\right)={\cal R}^{i}
\]
Let $\left(R_{0},R'_{0},R_{1}\right)\in ext\left({\cal R}^{o}\right)$.
If $R_{0}=0$ for example, then in order to show that $\left(0,R'_{0},R_{1}\right)\in{\cal R}^{i}$
we will prove that ${\cal R}^{o}$ is tight on the intersection of
${\cal R}^{o}$ with the hyper-plane $R_{0}=0$, and similarly for
$R'_{0}=0$ and $R_{1}=0$. We expect to get 2D capacity regions,
which correspond to the special cases presented in Section III after
the definition of ${\cal R}^{i}$.
\begin{lem}
\label{lem:Intersections Lemma}Let
\[
{\cal R}^{o}\left(R_{0}=0\right)\triangleq\left\{ \left(R'_{0},R{}_{1}\right)\vert\left(0,R'_{0},R{}_{1}\right)\in{\cal R}^{o}\right\} 
\]
Define ${\cal R}^{o}\left(R'_{0}=0\right)$ and ${\cal R}^{o}\left(R_{1}=0\right)$
similarly. Then
\begin{eqnarray*}
{\cal R}^{o}\left(R_{0}=0\right) & = & {\cal C}_{BCC}\\
{\cal R}^{o}\left(R'_{0}=0\right) & = & {\cal C}_{BC}\\
{\cal R}^{o}\left(R_{1}=0\right) & = & {\cal C}_{noR_{1}}
\end{eqnarray*}

\end{lem}
The proof of Lemma \ref{lem:Intersections Lemma} is given at the
end of this subsection.\\
We proceed to prove that $ext\left({\cal R}^{o}\right)\subseteq{\cal R}^{i}$.
Let $\left(R_{0},R'_{0},R_{1}\right)\in ext\left({\cal R}^{o}\right)$.
If $R_{0}=0$, then from the fact that ${\cal C}_{BC}$ is a special
case of ${\cal R}^{i}$, and from Lemma \ref{lem:Intersections Lemma}
above we can state
\[
If\,\,\,\left(0,R'_{0},R_{1}\right)\in ext\left({\cal R}^{o}\right)\Rightarrow\left(0,R'_{0},R_{1}\right)\in{\cal R}^{i}
\]
If $R'_{0}=0$ or $R_{1}=0$ we can do the same. The remaining extreme
points to be treated are in the positive orthant 
\[
\mathbb{R}_{++}^{3}\triangleq\left\{ \left(R_{0},R'_{0},R_{1}\right)\in\mathbb{R}^{k}\vert R_{0},R'_{0},R_{1}>0\right\} 
\]
To find them all we will examine an arbitrary region in ${\cal R}^{o}$.
Let $\left(U,V,X\right)\sim p\left(u,v,x\right)$ be any distribution.
It defines a convex bounded ${\cal H}$-polytope - an intersection
of eight half-spaces defined by the inequalities  \eqref{eq:UB}.
By Lemma \ref{lem:polytopes representation}, it has an equivalent
representation as a ${\cal V}$-polytope of it's own vertices, and
those vertices are the only candidates as extreme points of ${\cal R}^{o}$.
 Each vertex is obtained by only three (out of eight) linearly independent
active inequalities, as three linearly independent equations define
a specific point in $\mathbb{R}_{++}^{3}$. If there are less than
three independent active inequalities, we can have a straight line
(two active inequalities), a hyper-plane (one active inequality) or
interior point (no active inequalities) - but not a vertex. We are
interested only in the vertices in $\mathbb{R}_{++}^{3}$ if there
exist one - so inequalities (\ref{eq:UB6}) are in-active. Note that
if there are no vertices in $\mathbb{R}_{++}^{3}$, then all the vertices
has already been treated and proved to be in ${\cal R}^{i}$. The
three inequalities (\ref{eq:UB3})-(\ref{eq:UB5}) are all dependent,
thus only one of them is active (unless there is a redundancy, but
it does not affect the proof) thus inequalities (\ref{eq:UB1}) and
(\ref{eq:UB2}) are surely active. Thus, for such a vertex, we have:\begin{subequations} \label{eq:vertex_rep}
\begin{align}
R_{0}= & I\left(U;Y_{2}\right)\label{eq:vertex_rep1}\\
R_{0}+R'_{0}= & I\left(UV;Y_{2}\right)+C_{1}\label{eq:vertex_rep2}\\
R_{0}+R'_{0}+R_{1}\le & I\left(X;Y_{1}\right)\label{eq:vertex_rep3}\\
R_{0}+R'_{0}+R_{1}\le & I\left(U;Y_{2}\right)+I\left(X;Y_{1}\vert U\right)\label{eq:vertex_rep4}\\
R_{0}+R'_{0}+R_{1}\le & I\left(UV;Y_{2}\right)+C_{1}+I\left(X;Y_{1}\vert UV\right)\label{eq:vertex_rep5}
\end{align}
\end{subequations} where either (\ref{eq:vertex_rep3}), (\ref{eq:vertex_rep4})
or (\ref{eq:vertex_rep5}) is active. Subtract (\ref{eq:vertex_rep2})
from (\ref{eq:vertex_rep5}), and subtract (\ref{eq:vertex_rep1})
from (\ref{eq:vertex_rep2}),(\ref{eq:vertex_rep4}) to get an alternative
representation of the same vertex:\begin{subequations} \label{eq:vertex_alt_rep}
\begin{align}
R_{0}= & I\left(U;Y_{2}\right)\label{eq:vertex_alt_rep1}\\
R'_{0}= & I\left(V;Y_{2}\vert U\right)+C_{1}\label{eq:vertex_alt_rep2}\\
R_{1}\le & I\left(X;Y_{1}\vert UV\right)\label{eq:vertex_alt_rep3}\\
R'_{0}+R_{1}\le & I\left(X;Y_{1}\vert U\right)\label{eq:vertex_alt_rep4}\\
R_{0}+R'_{0}+R_{1}\le & I\left(X;Y_{1}\right)\label{eq:vertex_alt_rep5}
\end{align}
\end{subequations} where either (\ref{eq:vertex_alt_rep3}), (\ref{eq:vertex_alt_rep4})
or (\ref{eq:vertex_alt_rep5}) is active. This alternative representation
of the vertex obeys  \eqref{eq:IB}, implying that this vertex is
a point in the polytope induced from the same distribution $p\left(u,v,x\right)$
in ${\cal R}^{i}$. Thus for every polytope in ${\cal R}^{o}$, all
vertices in $\mathbb{R}_{++}^{3}$ were proved to be in ${\cal R}^{i}$,
implying that $ext\left({\cal R}^{o}\right)\subseteq{\cal R}^{i}$.
This complete the equivalence proof of the regions ${\cal R}^{o}$=${\cal R}^{i}$.\emph{}\\
\emph{Proof of Lemma \ref{lem:Intersections Lemma}}. We will show
the proof for ${\cal R}^{o}\left(R'_{0}=0\right)={\cal C}_{BC}$.
The others are treated the same and thus omitted. We have shown that
${\cal C}_{BC}$ is a special case of ${\cal R}^{i}$, and obviously
${\cal R}^{i}\subseteq{\cal R}^{o}$ - thus we have ${\cal C}_{BC}\subseteq{\cal R}^{o}\left(R'_{0}=0\right)$.
To show that ${\cal R}^{o}\left(R'_{0}=0\right)\subseteq{\cal C}_{BC}$,
consider ${\cal R}^{o}\left(R'_{0}=0\right)$. It contains all rates
$\left(R_{0},R_{1}\right)$ satisfying:\begin{subequations}
\begin{align}
R_{0}\le & I\left(U;Y_{2}\right)\label{eq:IntersectionsLemma_01}\\
R_{0}+R_{1}\le & I\left(X;Y_{1}\right)\label{eq:IntersectionsLemma_02}\\
R_{0}+R_{1}\le & I\left(U;Y_{2}\right)+I\left(X;Y_{1}\vert U\right)\label{eq:IntersectionsLemma_03}\\
R_{0}+R_{1}\le & I\left(UV;Y_{2}\right)+C_{1}+I\left(X;Y_{1}\vert UV\right)\label{eq:IntersectionsLemma_04}
\end{align}
\end{subequations}for some joint distribution $p\left(u,v,x\right)p\left(y_{1},y_{2}\vert x\right)$.
Let $\left(U,V,X\right)\sim p^{*}\left(u,v,x\right)$ be any distribution
that defines a 2D-polytope denoted by $P_{1}$. Our goal is to prove
this 2D-polytope is also in ${\cal C}_{BC}$. Recall ${\cal C}_{BC}$
- it contains all rates $\left(R_{0},R_{1}\right)$ satisfying:
\begin{align*}
R_{0}\le & I\left(U;Y_{2}\right)\\
R_{0}+R_{1}\le & I\left(X;Y_{1}\right)\\
R_{0}+R_{1}\le & I\left(U;Y_{2}\right)+I\left(X;Y_{1}\vert U\right)
\end{align*}
for some joint distribution $p\left(u,x\right)p\left(y_{1},y_{2}\vert x\right)$.\\
Consider the 2D-polytope $P_{2}$ contained in ${\cal C}_{BC}$, defined
by $\left(U,X\right)\sim p^{*}\left(u,x\right)$. We have that $P_{1}\subseteq P_{2}$
as both of them obey the same three inequalities (\ref{eq:IntersectionsLemma_01})-(\ref{eq:IntersectionsLemma_03}),
and that $P_{1}$ has one more inequality (\ref{eq:IntersectionsLemma_04})
to hold. It is true to any $P_{1}$ in ${\cal R}^{o}\left(R'_{0}=0\right),$
thus ${\cal R}^{o}\left(R'_{0}=0\right)\subseteq{\cal C}_{BC}$.

\subsection{Converse Part}

\begin{flushleft}
Let $\left(2^{nR_{0}},2^{nR'_{0}},2^{nR_{1}},n\right)$ be any sequence
of codes for the DM-BC with degraded message sets and unreliable conference
link that satisfies
\begin{eqnarray*}
\underset{{\scriptscriptstyle n\longrightarrow\infty}}{\lim}P_{e}^{\left(n\right)}=0 & \,\,\,\,\, & \underset{{\scriptscriptstyle n\longrightarrow\infty}}{\lim}P_{e}^{'\left(n\right)}=0
\end{eqnarray*}
We have to show that the inequalities  \eqref{eq:UB} hold, for some
pmf $p\left(u,v,x\right)$. Applying the Fano's inequality:
\begin{flalign}
 & n\left(R_{0}-\epsilon{}_{2,n}\right)\le I\left(M_{0};Y_{2}^{n}\right)\nonumber \\
 & =\stackrel[{\scriptstyle i=1}]{{\scriptstyle n}}{\sum}I\left(M_{0};Y_{2,i}\vert Y_{2,i+1}^{\,\,n}\right)\nonumber \\
 & \le\stackrel[{\scriptstyle i=1}]{{\scriptstyle n}}{\sum}I\left(M_{0}Y_{2,i+1}^{\,\,n}Y_{1}^{i-1};Y_{2,i}\right)\label{eq:UBproof1}\\
\nonumber \\
 & n\left(R_{0}+R'_{0}-\epsilon'{}_{2,n}\right)\le I\left(M_{0}M'_{0};Y_{2}^{n},\phi\left(Y_{1}^{n}\right)\right)\nonumber \\
 & =I\left(M_{0}M'_{0};Y_{2}^{n}\right)+I\left(M_{0}M'_{0};\phi\left(Y_{1}^{n}\right)\vert Y_{2}^{n}\right)\nonumber \\
 & \le\stackrel[{\scriptstyle i=1}]{{\scriptstyle n}}{\sum}I\left(M_{0}M'_{0};Y_{2,i}\vert Y_{2,i+1}^{\,\,n}\right)+H\left(\phi\left(Y_{1}^{n}\right)\right)\nonumber \\
 & \le\stackrel[{\scriptstyle i=1}]{{\scriptstyle n}}{\sum}I\left(M_{0}M'_{0}Y_{2,i+1}^{\,\,n};Y_{2,i}\right)+nC_{1}\nonumber \\
 & \le\stackrel[{\scriptstyle i=1}]{{\scriptstyle n}}{\sum}I\left(M_{0}M'_{0}Y_{2,i+1}^{\,\,n}Y_{1}^{i-1};Y_{2,i}\right)+nC_{1}\label{eq:UBproof2}\\
\nonumber \\
 & n\left(R_{0}+R'_{0}+R_{1}-\epsilon{}_{1,n}\right)\le I\left(M_{0}M'_{0}M_{1};Y_{1}^{n}\right)\nonumber \\
 & =\stackrel[{\scriptstyle i=1}]{{\scriptstyle n}}{\sum}I\left(M_{0}M'_{0}M_{1};Y_{1,i}\vert Y_{1}^{i-1}\right)\nonumber \\
 & \le\stackrel[{\scriptstyle i=1}]{{\scriptstyle n}}{\sum}I\left(X_{i}M_{0}M'_{0}M_{1}Y_{1}^{i-1};Y_{1,i}\right)\nonumber \\
 & =\stackrel[{\scriptstyle i=1}]{{\scriptstyle n}}{\sum}I\left(X_{i};Y_{1,i}\right)\label{eq:UBproof3}
\end{flalign}
where $\epsilon{}_{1,n}$,$\epsilon{}_{2,n}$,$\epsilon'{}_{2,n}$
tend to zero as $n\rightarrow\infty$. To prove inequalities (\ref{eq:UB4})
and (\ref{eq:UB5}) we bound:
\begin{align}
 & n\left(R_{0}+R'_{0}+R_{1}-\epsilon'{}_{2,n}-\epsilon{}_{1,n}\right)\nonumber \\
 & \le I\left(M_{0};Y_{2}^{n}\right)+I\left(M'_{0}M_{1};Y_{1}^{n}\vert M_{0}\right)\nonumber \\
 & =\stackrel[{\scriptstyle i=1}]{{\scriptstyle n}}{\sum}I\left(M_{0};Y_{2,i}\vert Y_{2,i+1}^{\,\,n}\right)+\stackrel[{\scriptstyle i=1}]{{\scriptstyle n}}{\sum}I\left(M'_{0}M{}_{1};Y_{1,i}\vert M_{0}Y_{1}^{i-1}\right)\nonumber \\
 & \le\stackrel[{\scriptstyle i=1}]{{\scriptstyle n}}{\sum}I\left(M{}_{0}Y_{2,i+1}^{\,\,n};Y_{2,i}\right)\nonumber \\
 & +\stackrel[{\scriptstyle i=1}]{{\scriptstyle n}}{\sum}I\left(M'_{0}M{}_{1}Y_{2,i+1}^{\,\,n};Y_{1,i}\vert M_{0}Y_{1}^{i-1}\right)\nonumber \\
 & =\stackrel[{\scriptstyle i=1}]{{\scriptstyle n}}{\sum}I\left(M{}_{0}Y_{2,i+1}^{\,\,n};Y_{2,i}\right)+\stackrel[{\scriptstyle i=1}]{{\scriptstyle n}}{\sum}I\left(Y_{2,i+1}^{\,\,n};Y_{1,i}\vert M_{0}Y_{1}^{i-1}\right)\nonumber \\
 & +\stackrel[{\scriptstyle i=1}]{{\scriptstyle n}}{\sum}I\left(M'_{0}M{}_{1};Y_{1,i}\vert M_{0}Y_{1}^{i-1}Y_{2,i+1}^{\,\,n}\right)\nonumber \\
 & \overset{\left(a\right)}{=}\stackrel[{\scriptstyle i=1}]{{\scriptstyle n}}{\sum}I\left(M{}_{0}Y_{2,i+1}^{\,\,n};Y_{2,i}\right)+\stackrel[{\scriptstyle i=1}]{{\scriptstyle n}}{\sum}I\left(Y_{1}^{i-1};Y_{2,i}\vert M_{0}Y_{2,i+1}^{\,\,n}\right)\nonumber \\
 & +\stackrel[{\scriptstyle i=1}]{{\scriptstyle n}}{\sum}I\left(X_{i}M'_{0}M{}_{1};Y_{1,i}\vert M_{0}Y_{1}^{i-1}Y_{2,i+1}^{\,\,n}\right)\nonumber \\
 & =\stackrel[{\scriptstyle i=1}]{{\scriptstyle n}}{\sum}I\left(M{}_{0}Y_{2,i+1}^{\,\,n}Y_{1}^{i-1};Y_{2,i}\right)\nonumber \\
 & +\stackrel[{\scriptstyle i=1}]{{\scriptstyle n}}{\sum}I\left(X_{i};Y_{1,i}\vert M_{0}Y_{1}^{i-1}Y_{2,i+1}^{\,\,n}\right)\label{eq:UBproof4}\\
\nonumber \\
 & n\left(R_{0}+R'_{0}+R_{1}-\epsilon'{}_{2,n}-\epsilon{}_{1,n}\right)\nonumber \\
 & \le I\left(M_{0}M'_{0};Y_{2}^{n},\phi\left(Y_{1}^{n}\right)\right)+I\left(M_{1};Y_{1}^{n}\vert M_{0}M'_{0}\right)\nonumber \\
 & \overset{\left(b\right)}{\le}\stackrel[{\scriptstyle i=1}]{{\scriptstyle n}}{\sum}I\left(M_{0}M'_{0}Y_{2,i+1}^{\,\,n};Y_{2,i}\right)+nC_{1}\nonumber \\
 & +\stackrel[{\scriptstyle i=1}]{{\scriptstyle n}}{\sum}I\left(M{}_{1}Y_{2,i+1}^{\,\,n};Y_{1,i}\vert M_{0}M'_{0}Y_{1}^{i-1}\right)\nonumber \\
 & \le\stackrel[{\scriptstyle i=1}]{{\scriptstyle n}}{\sum}I\left(M_{0}M'_{0}Y_{2,i+1}^{\,\,n};Y_{2,i}\right)+nC_{1}\nonumber \\
 & +\stackrel[{\scriptstyle i=1}]{{\scriptstyle n}}{\sum}I\left(Y_{2,i+1}^{\,\,n};Y_{1,i}\vert M_{0}M'_{0}Y_{1}^{i-1}\right)\nonumber \\
 & +\stackrel[{\scriptstyle i=1}]{{\scriptstyle n}}{\sum}I\left(M{}_{1};Y_{1,i}\vert M_{0}M'_{0}Y_{1}^{i-1}Y_{2,i+1}^{\,\,n}\right)\nonumber \\
 & \overset{\left(a\right)}{=}\stackrel[{\scriptstyle i=1}]{{\scriptstyle n}}{\sum}I\left(M_{0}M'_{0}Y_{2,i+1}^{\,\,n};Y_{2,i}\right)+nC_{1}\nonumber \\
 & +\stackrel[{\scriptstyle i=1}]{{\scriptstyle n}}{\sum}I\left(Y_{1}^{i-1};Y_{2,i}\vert M_{0}M'_{0}Y_{2,i+1}^{\,\,n}\right)\nonumber \\
 & +\stackrel[{\scriptstyle i=1}]{{\scriptstyle n}}{\sum}I\left(X_{i}M{}_{1};Y_{1,i}\vert M_{0}M'_{0}Y_{1}^{i-1}Y_{2,i+1}^{\,\,n}\right)\nonumber \\
 & =\stackrel[{\scriptstyle i=1}]{{\scriptstyle n}}{\sum}I\left(M_{0}Y_{2,i+1}^{\,\,n}Y_{1}^{i-1}M'_{0};Y_{2,i}\right)+nC_{1}\nonumber \\
 & +\stackrel[{\scriptstyle i=1}]{{\scriptstyle n}}{\sum}I\left(X_{i};Y_{1,i}\vert M_{0}Y_{1}^{i-1}Y_{2,i+1}^{\,\,n}M'_{0}\right)\label{eq:UBproof5}
\end{align}
where $\left(a\right)$ is due to Csiszar sum identity and the fact
that $X_{i}$ is a deterministic function of the messages, and $\left(b\right)$
is from inequality (\ref{eq:UBproof2}) above. Finally back to inequalities
(\ref{eq:UBproof1})-(\ref{eq:UBproof5}) and define $U_{i}=\left(M_{0}Y_{2,i+1}^{\,\,n}Y_{1}^{i-1}\right)$,
$V_{i}=M'_{0}$. Then define also a time-sharing random variable uniformly
distributed $Q\sim Uniform\left[1:n\right]$ independent of all R.V,
and taking the limit $n\rightarrow\infty$ to complete the converse
proof.
\par\end{flushleft}

\subsection{Direct Part}

The encoder utilizes a superposition coding scheme and uses binning
for the residual message.

\subsubsection*{Codebook Generation}

Fix $P_{U}P_{V\vert U}P_{X\vert UV}$. Generate the codebook ${\cal C}$
as follows:
\begin{enumerate}
\item Generate $2^{nR_{0}}$ independent codewords 
\[
\boldsymbol{u}\left(m_{0}\right)\thicksim\prod_{i=1}^{n}P_{U}\left(u_{i}\right)
\]
where $m_{0}\in\left[1:2^{nR_{0}}\right]$.
\item For each codeword $\boldsymbol{u}\left(m_{0}\right)$, generate $2^{nR'_{0}}$
conditionally independent codewords 
\[
\boldsymbol{v}\left(m_{0},m'_{0}\right)\thicksim\prod_{i=1}^{n}P_{V\vert U}\left(v\vert u_{i}\left(m_{0}\right)\right)
\]
where $m'_{0}\in\left[1:2^{nR'_{0}}\right]$.
\item For each couple of codewords $\boldsymbol{u}\left(m_{0}\right),\boldsymbol{v}\left(m_{0},m'_{0}\right)$,
generate $2^{nR_{1}}$ conditionally independent codewords 
\[
\boldsymbol{x}\left(m_{0},m'_{0},m_{1}\right)\thicksim\prod_{i=1}^{n}P_{X\vert UV}\left(x_{i}\vert u_{i}\left(m_{0}\right),v_{i}\left(m_{0},m'_{0}\right)\right)
\]
where $m_{1}\in\left[1:2^{nR_{1}}\right]$.
\item Divide the residual message set $\left[1:2^{nR'_{0}}\right]$ into
$2^{nC_{1}}$ equal bins, each bin contains $2^{n\left(R'_{0}-C_{1}\right)}$
messages. Denote by $bin\left(m\right)$ the bin index of any message
$m\in\left[1:2^{nR'_{0}}\right]$.
\end{enumerate}
This defines the codebook: \\
\[
{\cal C}=\left\{ \begin{array}{c}
\left[\boldsymbol{u}\left(m_{0}\right),\boldsymbol{v}\left(m_{0},m'_{0}\right),\boldsymbol{x}\left(m_{0},m'_{0},m_{1}\right)\right]\\
\left(m_{0},m'_{0},m_{1}\right)\in\left[1:2^{nR_{0}}\right]\vartimes\left[1:2^{nR'_{0}}\right]\vartimes\left[1:2^{nR_{1}}\right]
\end{array}\right\} 
\]

\subsubsection*{Encoding and decoding scheme}

\textbf{Encoder:} Let $\left(m_{0},m'_{0},m_{1}\right)\in\left[1:2^{nR_{0}}\right]\vartimes\left[1:2^{nR'_{0}}\right]\vartimes\left[1:2^{nR_{1}}\right]$
be the messages to be sent. The encoder transmits $\boldsymbol{x}\left(m_{0},m'_{0},m_{1}\right)$
from codebook ${\cal C}$.\textbf{}\\
\textbf{Decoder 1} finds the unique triplet $\left(\hat{m}_{0},\hat{m}'_{0},\hat{m}_{1}\right)$
such that 
\[
\left(\boldsymbol{u}\left(\hat{m}_{0}\right),\boldsymbol{v}\left(\hat{m}_{0},\hat{m}'_{0}\right),\boldsymbol{x}\left(\hat{m}_{0},\hat{m}'_{0},\hat{m}_{1}\right),\boldsymbol{y}_{1}\right)\in{\cal T}_{e}^{\left(n\right)}
\]
By standard techniques, decoder 1 decodes $\left(m_{0},m'_{0},m_{1}\right)$\textbf{
}correctly, with an arbitrarily small probability of error, if\begin{subequations} \label{eq:IBproofDec1}
\begin{align}
R_{1}\le & I\left(X;Y_{1}\vert UV\right)\label{eq:IBproofDec1_1}\\
R'_{0}+R_{1}\le & I\left(X;Y_{1}\vert U\right)\label{eq:IBproofDec1_2}\\
R_{0}+R'_{0}+R_{1}\le & I\left(X;Y_{1}\right)\label{eq:IBproofDec1_3}
\end{align}
\end{subequations}In the end of the transmission and if the conference
link is present, decoder 1 sends to decoder 2 the estimated residual
message's bin index $bin\left(\hat{m}'_{0}\right)$.\textbf{}\\
\textbf{Decoder 2} - the operation of decoder 2 depends on whether
the conference link is present

If the \textbf{conference link is absent, decoder 2} finds the unique
message $\check{m}_{0}$ such that 
\[
\left(\boldsymbol{u}\left(\check{m}_{0}\right),\boldsymbol{y}_{2}\right)\in{\cal T}_{e}^{\left(n\right)}
\]
By standard techniques, decoder 2 decodes $m_{0}$\textbf{ }correctly,
with an arbitrarily small probability of error, if\begin{subequations}
\begin{equation}
R_{0}\le I\left(U;Y_{2}\right)\label{eq:R0<I(U;Y2)}
\end{equation}

If the \textbf{conference link is present, decoder 2} finds the unique
couple of messages $\left(\check{m}_{0},\check{m}'_{0}\right)$ such
that 
\[
\left(\boldsymbol{u}\left(\check{m}_{0}\right),\boldsymbol{v}\left(\check{m}_{0},\check{m}'_{0}\right),\boldsymbol{y}_{2}\right)\in{\cal T}_{e}^{\left(n\right)}
\]
and where $bin\left(\check{m}'_{0}\right)=\phi\left(\boldsymbol{y}_{1}\right)$.\\
By standard techniques, decoder 2 decodes $\left(m_{0},m'_{0}\right)$\textbf{
}correctly, with an arbitrarily small probability of error, if 
\begin{align}
R'_{0}-C_{1}\le & I\left(V;Y_{2}\vert U\right)\label{eq:R'0-C1<I(V;Y2|U)}\\
R{}_{0}+R'_{0}-C_{1}\le & I\left(UV;Y_{2}\right)\label{eq:R0+R'0-C1<I(UV;Y2)}
\end{align}
\end{subequations}Note that inequality (\ref{eq:R0+R'0-C1<I(UV;Y2)})
is already satisfied, as inequalities (\ref{eq:R0<I(U;Y2)}) and (\ref{eq:R'0-C1<I(V;Y2|U)})
are satisfied. The direct part follows by  \eqref{eq:IBproofDec1},
(\ref{eq:R0<I(U;Y2)}) and (\ref{eq:R'0-C1<I(V;Y2|U)}).

\subsubsection*{Error Probability Analysis}

Assume without loss of generality, that $\left(m_{0},m'_{0},m_{1}\right)=\left(1,1,1\right)$
were sent, and that $bin\left(m'_{0}\right)=1$.\textbf{ }Define the
error event ${\cal E}_{1}$: 
\begin{eqnarray*}
{\cal E}_{1} & = & \left\{ \left(\hat{M}_{0},\hat{M}'_{0},\hat{M}_{1}\right)\neq\left(1,1,1\right)\right\} 
\end{eqnarray*}
\textbf{Decoder 1} makes an error only if one or more of the following
events occur:
\begin{eqnarray*}
{\cal E}{}_{10} & = & \left\{ \left(\boldsymbol{u}\left(1\right),\boldsymbol{v}\left(1,1\right),\boldsymbol{x}\left(1,1,1\right),\boldsymbol{y}_{1}\right)\notin{\cal T}_{e}^{\left(n\right)}\right\} \\
{\cal E}{}_{11} & = & \underset{{\scriptscriptstyle \left(m_{0},m'_{0},m_{1}\right)\neq\left(1,1,1\right)}}{{\displaystyle \bigcup}}{\cal E}{\scriptstyle \left(m_{0},m'_{0},m_{1}\right)}
\end{eqnarray*}
\begin{alignat*}{1}
{\cal E}{\scriptstyle \left(m_{0},m'_{0},m_{1}\right)}=\left\{ \left(\boldsymbol{u}{\scriptstyle \left(m_{0}\right)},\boldsymbol{v}{\scriptstyle \left(m_{0},m'_{0}\right)},\boldsymbol{x}{\scriptstyle \left(m_{0},m'_{0},m_{1}\right)},\boldsymbol{y}_{1}\right)\in{\cal T}_{e}^{\left(n\right)}\right\} 
\end{alignat*}
where the error events ${\cal E}{\scriptstyle \left(m_{0},m'_{0},m_{1}\right)}$
are divided into eight different groups, each group contains exponentially
many error events of the same kind. The probability of ${\cal E}_{10}$
tends to zero as $n\rightarrow\infty$ by LLN, and the probability
of the other error events tend to zero by the Packing Lemma \cite{ElGamal2011}
if:\begin{subequations} \label{eq:IBproof}
\begin{align}
R_{1}\le & I\left(X;Y_{1}\vert UV\right)\label{eq:IBproof1}\\
R'_{0}\le & I\left(X;Y_{1}\vert U\right)\label{eq:IBproof2}\\
R'_{0}+R_{1}\le & I\left(X;Y_{1}\vert U\right)\label{eq:IBproof3}\\
R_{0}\le & I\left(X;Y_{1}\right)\label{eq:IBproof4}\\
R_{0}+R_{1}\le & I\left(X;Y_{1}\right)\label{eq:IBproof5}\\
R_{0}+R'_{0}\le & I\left(X;Y_{1}\right)\label{eq:IBproof6}\\
R_{0}+R'_{0}+R_{1}\le & I\left(X;Y_{1}\right)\label{eq:IBproof7}
\end{align}
\end{subequations}Note that (\ref{eq:IBproof2}) and (\ref{eq:IBproof4})-(\ref{eq:IBproof6})
are redundant as (\ref{eq:IBproof3}) and (\ref{eq:IBproof7}) hold,
respectively. The joint pmf of each group and the relevant rates constraint
has to be satisfied in order to have $P\left({\cal E}\right)\rightarrow0$,
is depicted in Table (\ref{tab:ErrorEvents}).
\begin{table}
\begin{centering}
\begin{tabular}{|c|c|c|c|c|c|}
\hline 
 ${\scriptstyle {\cal E}}$ & ${\scriptstyle m_{0}}$ & ${\scriptstyle m'_{0}}$ & ${\scriptstyle m_{1}}$ & ${\scriptstyle p\left(\boldsymbol{u},\boldsymbol{v},\boldsymbol{x},\boldsymbol{y}_{1}\right)}$ & ${\scriptstyle P\left({\cal E}\right)\rightarrow0}$\tabularnewline
\hline 
\hline 
${\cal E}{}_{10}$ & 1 & 1 & 1 & ${\scriptstyle p\left(\boldsymbol{u},\boldsymbol{v},\boldsymbol{x}\right)p\left(\boldsymbol{y}_{1}\vert\boldsymbol{x}\right)}$ & ---\tabularnewline
\hline 
${\cal E}{\scriptstyle \left(1,1,*\right)}$ & 1 & 1 & {*} & ${\scriptstyle p\left(\boldsymbol{u},\boldsymbol{v},\boldsymbol{x}\right)p\left(\boldsymbol{y}_{1}\vert\boldsymbol{u},\boldsymbol{v}\right)}$ & (\ref{eq:IBproof1})\tabularnewline
\hline 
${\cal E}{\scriptstyle \left(1,*,1\right)}$ & 1 & {*} & 1 & ${\scriptstyle p\left(\boldsymbol{u},\boldsymbol{v},\boldsymbol{x}\right)p\left(\boldsymbol{y}_{1}\vert\boldsymbol{u}\right)}$ & (\ref{eq:IBproof2})\tabularnewline
\hline 
${\cal E}{\scriptstyle \left(1,*,*\right)}$ & 1 & {*} & {*} & ${\scriptstyle p\left(\boldsymbol{u},\boldsymbol{v},\boldsymbol{x}\right)p\left(\boldsymbol{y}_{1}\vert\boldsymbol{u}\right)}$ & (\ref{eq:IBproof3})\tabularnewline
\hline 
${\cal E}{\scriptstyle \left(*,1,1\right)}$ & {*} & 1 & 1 & ${\scriptstyle p\left(\boldsymbol{u},\boldsymbol{v},\boldsymbol{x}\right)p\left(\boldsymbol{y}_{1}\right)}$ & (\ref{eq:IBproof4})\tabularnewline
\hline 
${\cal E}{\scriptstyle \left(*,1,*\right)}$ & {*} & 1 & {*} & ${\scriptstyle p\left(\boldsymbol{u},\boldsymbol{v},\boldsymbol{x}\right)p\left(\boldsymbol{y}_{1}\right)}$ & (\ref{eq:IBproof5})\tabularnewline
\hline 
${\cal E}{\scriptstyle \left(*,*,1\right)}$ & {*} & {*} & 1 & ${\scriptstyle p\left(\boldsymbol{u},\boldsymbol{v},\boldsymbol{x}\right)p\left(\boldsymbol{y}_{1}\right)}$ & (\ref{eq:IBproof6})\tabularnewline
\hline 
${\cal E}{\scriptstyle \left(*,*,*\right)}$ & {*} & {*} & {*} & ${\scriptstyle p\left(\boldsymbol{u},\boldsymbol{v},\boldsymbol{x}\right)p\left(\boldsymbol{y}_{1}\right)}$ & (\ref{eq:IBproof7})\tabularnewline
\hline 
\end{tabular}
\par\end{centering}

\protect\caption{\label{tab:ErrorEvents}The error events of decoder 1}
\end{table}
The probability of error event ${\cal E}{}_{1}$ is upper bounded
as:
\begin{eqnarray*}
P\left({\cal E}_{1}\right) & \triangleq & P\left\{ \left(\hat{M}_{0},\hat{M}'_{0},\hat{M}_{1}\right)\neq\left(1,1,1\right)\right\} \\
 & = & P\left({\cal E}{}_{10}\cup{\cal E}{}_{11}\right)\le P\left({\cal E}{}_{10}\right)+P\left({\cal E}{}_{11}\right)
\end{eqnarray*}
the first term tends to zero as $n\rightarrow\infty$ by LLN, and
the second term tends to zero as $n\rightarrow\infty$ if (\ref{eq:IBproof1}),(\ref{eq:IBproof3})
and (\ref{eq:IBproof7}) hold.\\
Define the error event ${\cal E}_{2}$:
\begin{eqnarray*}
{\cal E}_{2} & = & \left\{ \check{M}_{0}\neq1\right\} 
\end{eqnarray*}
If the conference link is absent, \textbf{decoder 2} makes an error
only if one or more of the following events occur:
\begin{eqnarray*}
{\cal E}{}_{20} & = & \left\{ \left(\boldsymbol{u}\left(1\right),\boldsymbol{y}_{2}\right)\notin{\cal T}_{e}^{\left(n\right)}\right\} \\
{\cal E}{}_{21} & = & \left\{ \left(\boldsymbol{u}{\scriptstyle \left(m_{0}\right)},\boldsymbol{y}_{2}\right)\in{\cal T}_{e}^{\left(n\right)}\,s.t.\,{\scriptstyle m_{0}\neq1}\right\} 
\end{eqnarray*}
Following the same analysis, the probability of error event ${\cal E}{}_{2}$
is upper bounded as:
\begin{eqnarray*}
P\left({\cal E}_{2}\right) & \triangleq & P\left\{ \check{M}_{0}\neq1\right\} \\
 & = & P\left({\cal E}{}_{20}\cup{\cal E}{}_{21}\right)\le P\left({\cal E}{}_{20}\right)+P\left({\cal E}{}_{21}\right)
\end{eqnarray*}
the first term tends to zero as $n\rightarrow\infty$ by LLN, and
by Packing Lemma the second term tends to zero as $n\rightarrow\infty$
if
\begin{equation}
R_{0}\le I\left(U;Y_{2}\right)\label{eq:IBproof2_3}
\end{equation}
Define the error event ${\cal E}'_{2}$:
\begin{eqnarray*}
{\cal E}'_{2} & = & \left\{ \left(\check{M}_{0},\check{M}'_{0}\right)\neq\left(1,1\right)\right\} 
\end{eqnarray*}
If the conference link is present, \textbf{decoder 2} makes an error
only if one or more of the following events occur:
\begin{eqnarray*}
{\cal E}'{}_{20} & = & \left\{ \left(\boldsymbol{u}\left(1\right),\boldsymbol{v}\left(1,1\right),\boldsymbol{y}_{2}\right)\notin{\cal T}_{e}^{\left(n\right)}\right\} \\
{\cal E}'{}_{21} & = & \{\left(\boldsymbol{u}{\scriptstyle \left(m_{0}\right)},\boldsymbol{v}{\scriptstyle \left(m_{0},m'_{0}\right)},\boldsymbol{y}_{2}\right)\in{\cal T}_{e}^{\left(n\right)}\\
 &  & \,s.t.\,{\scriptstyle \left(m_{0},m'_{0}\right)}\neq{\scriptstyle \left(1,1\right)},\,{\scriptstyle bin\left(m'_{0}\right)}={\scriptstyle 1}\}
\end{eqnarray*}
The probability of error event ${\cal E}'{}_{2}\cap{\cal E}_{1}^{C}$
is upper bounded as:
\begin{eqnarray*}
P\left({\cal E}'_{2}\cap{\cal E}_{1}^{C}\right) & = & P\left(\left\{ \left(\check{M}_{0},\check{M}'_{0}\right)\neq\left(1,1\right)\right\} \cap{\cal E}_{1}^{C}\right)\\
 & = & P\left(\left({\cal E}'{}_{20}\cup{\cal E}'{}_{21}\right)\cap{\cal E}_{1}^{C}\right)\\
 & \le & P\left({\cal E}'{}_{20}\right)+P\left({\cal E}'{}_{21}\cap{\cal E}_{1}^{C}\right)
\end{eqnarray*}
the first term tends to zero as $n\rightarrow\infty$ by LLN, and
by Packing Lemma the second term tends to zero as $n\rightarrow\infty$
if\begin{subequations}
\begin{eqnarray}
R'_{0}-C_{1} & \le & I\left(V;Y_{2}\vert U\right)\label{eq:IBproof2_1}\\
R_{0}+R'_{0}-C_{1} & \le & I\left(UV;Y_{2}\right)\label{eq:IBproof2_2}
\end{eqnarray}
\end{subequations}Note that (\ref{eq:IBproof2_2}) is redundant as
(\ref{eq:IBproof2_1}) and (\ref{eq:IBproof2_3}) hold. Finally $P_{e}^{\left(n\right)}$
and $P_{e}^{'\left(n\right)}$ are upper bounded as:
\begin{eqnarray*}
P_{e}^{\left(n\right)} & = & P\left({\cal E}{}_{1}\cup{\cal E}{}_{2}\right)\le P\left({\cal E}{}_{1}\right)+P\left({\cal E}{}_{2}\right)\\
P_{e}^{'\left(n\right)} & = & P\left({\cal E}{}_{1}\cup{\cal E}'{}_{2}\right)=P\left({\cal E}{}_{1}\right)+P\left({\cal E}'{}_{2}\cap{\cal E}_{1}^{C}\right)
\end{eqnarray*}
where both $P_{e}^{\left(n\right)}$ and $P_{e}^{'\left(n\right)}$
tend to zero if  \eqref{eq:UB} hold. This complete the direct part.

\subsection{Proof for Case 3}

\begin{figure}
\includegraphics[scale=0.7]{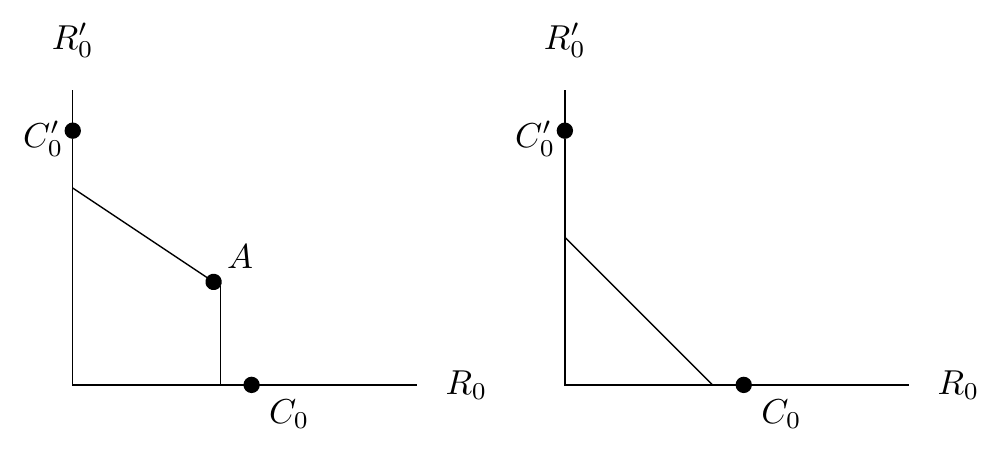}

\protect\caption{\label{fig:The-region-shapes}The optional region shapes in ${\cal R}_{noR_{1}}^{o}$}
\end{figure}

Let ${\cal R}_{noR_{1}}^{i}$ be the set of all rate pairs $\left(R_{0},R'_{0}\right)$
satisfying:
\begin{align*}
R_{0}\le & I\left(U;Y_{2}\right)\\
R'_{0}\le & min\left\{ I\left(X;Y_{1}\vert U\right),I\left(X;Y_{2}\vert U\right)+C_{1}\right\} \\
R_{0}+R'_{0}\le & I\left(X;Y_{1}\right)
\end{align*}
for some joint distribution $p\left(u,x\right)p\left(y_{1},y_{2}\vert x\right)$.
It is the same achievable region defined in Section III Case 3. Until
proven otherwise, we shall treat this region only as an achievable
region. Let ${\cal R}_{noR_{1}}^{o}$ be the set of all rate pairs
$\left(R_{0},R'_{0}\right)$ satisfying:\begin{subequations} \label{eq:UBcommon}
\begin{align}
R_{0}\le & I\left(U;Y_{2}\right)\label{eq:UBcommon1}\\
R_{0}+R'_{0}\le & I\left(U;Y_{2}\right)+I\left(X;Y_{1}\vert U\right)\label{eq:UBcommon2}\\
R_{0}+R'_{0}\le & I\left(X;Y_{2}\right)+C_{1}\label{eq:UBcommon3}\\
R_{0}+R'_{0}\le & I\left(X;Y_{1}\right)\label{eq:UBcommon4}
\end{align}
\end{subequations} for some joint distribution $p\left(u,x\right)p\left(y_{1},y_{2}\vert x\right)$. 
\begin{thm}
The capacity region for the DM-BC with unreliable conference link
and common messages only is given by
\[
{\cal C}_{noR_{1}}={\cal R}_{noR_{1}}^{o}={\cal R}_{noR_{1}}^{i}
\]

\end{thm}
The proof of the capacity region for this 2D model stimulated the
proof for the general model defined in Section II. The region ${\cal R}_{noR_{1}}^{i}$
is achievable as a special case of ${\cal R}^{i}$, see Section III
Case 3. We proceed by proving that ${\cal R}_{noR_{1}}^{o}={\cal R}_{noR_{1}}^{i}$
and then prove the converse part, i.e. ${\cal C}_{noR_{1}}\subseteq{\cal R}_{noR_{1}}^{o}$,
which concludes that ${\cal C}_{noR_{1}}={\cal R}_{noR_{1}}^{o}={\cal R}_{noR_{1}}^{i}$.
It is straight forward to prove that ${\cal R}_{noR_{1}}^{i}\subseteq{\cal R}_{noR_{1}}^{o}$
- for every $p\left(u,x\right)$, the inequalities satisfied in ${\cal R}_{noR_{1}}^{i}$
imply that the inequalities in ${\cal R}_{noR_{1}}^{o}$ are also
satisfied. In order to prove ${\cal R}_{noR_{1}}^{o}\subseteq{\cal R}_{noR_{1}}^{i}$
we cannot use the last argument. Instead we will examine the corner
points of an arbitrary region in ${\cal R}_{noR_{1}}^{o}$. Define
the capacities:
\begin{eqnarray*}
C_{0} & \triangleq & \underset{p\left(x\right)}{\max}\min\left\{ I\left(X;Y_{1}\right),\left(X;Y_{2}\right)\right\} \\
C'_{0} & \triangleq & \underset{p\left(x\right)}{\max}\min\left\{ I\left(X;Y_{1}\right),\left(X;Y_{2}\right)+C_{1}\right\} 
\end{eqnarray*}
The line segments from the origin to $\left(C_{0},0\right)$ and
$\left(0,C'_{0}\right)$ are in ${\cal R}_{noR_{1}}^{o}$ and ${\cal R}_{noR_{1}}^{i}$.
The remaining points to examine are $\left\{ \left(R_{0},R'_{0}\right)\in{\cal R}_{noR_{1}}^{o}:R_{0},R'_{0}>0\right\} $.
Observe that for every $p\left(u,x\right)$, inequalities  \eqref{eq:UBcommon}
define a triangle or a trapezoid. The only points of interest are
the corner points like the point A in Fig. \ref{fig:The-region-shapes},
as if we prove that a corner point is in ${\cal R}_{noR_{1}}^{i}$,
we can say that all the other points in the trapezoid are also in
${\cal R}_{noR_{1}}^{i}$ using the convexity property. Note that
in the triangle shape, there are no points of interest because the
triangle is surely contained in ${\cal R}_{noR_{1}}^{i}$, as the
triangle's hypotenuse edges intersect the axes on the line segments
contained in ${\cal R}_{noR_{1}}^{i}$ and due to the convexity property.
Each corner point A is the intersection of two linearly independent
active constraints:\begin{subequations} \label{eq:UBcommonR}
\begin{align}
R_{0}= & I\left(U;Y_{2}\right)\label{eq:UBcommonR1}\\
R_{0}+R'_{0}\le & I\left(U;Y_{2}\right)+I\left(X;Y_{1}\vert U\right)\label{eq:UBcommonR2}\\
R_{0}+R'_{0}\le & I\left(X;Y_{2}\right)+C_{1}\label{eq:UBcommonR3}\\
R_{0}+R'_{0}\le & I\left(X;Y_{1}\right)\label{eq:UBcommonR4}
\end{align}
\end{subequations} where (\ref{eq:UBcommonR1}) and one of the three
(\ref{eq:UBcommonR2})-(\ref{eq:UBcommonR4}) are active. Subtract
(\ref{eq:UBcommonR1}) from (\ref{eq:UBcommonR2}) and (\ref{eq:UBcommonR3})
to have that A is also ${\cal R}_{noR_{1}}^{o}$ since inequalities
 \eqref{eq:UBcommon} hold. This complete the equivalence proof of
the regions ${\cal R}_{noR_{1}}^{o}$ and ${\cal R}_{noR_{1}}^{i}$.

The proof that ${\cal R}_{noR_{1}}^{o}$ is an upper bound, i.e. ${\cal C}_{noR_{1}}\subseteq{\cal R}_{noR_{1}}^{o}$,
is similar to the converse part given in Section IV.B. Thus we will
give it here shortly:\begin{subequations} \label{eq:UBcommonProof}
\begin{flalign}
 & n\left(R_{0}-\epsilon{}_{2,n}\right)\le I\left(M_{0};Y_{2}^{n}\right)\nonumber \\
 & =\stackrel[{\scriptstyle i=1}]{{\scriptstyle n}}{\sum}I\left(M_{0};Y_{2,i}\vert Y_{2,i+1}^{\,\,n}\right)\nonumber \\
 & \le\stackrel[{\scriptstyle i=1}]{{\scriptstyle n}}{\sum}I\left(M_{0}Y_{2,i+1}^{\,\,n}Y_{1}^{i-1};Y_{2,i}\right)\\
\nonumber \\
 & n\left(R_{0}+R'_{0}-\epsilon'{}_{2,n}\right)\le I\left(M_{0}M'_{0};Y_{2}^{n},\phi\left(Y_{1}^{n}\right)\right)\nonumber \\
 & =I\left(M_{0}M'_{0};Y_{2}^{n}\right)+I\left(M_{0}M'_{0};\phi\left(Y_{1}^{n}\right)\vert Y_{2}^{n}\right)\nonumber \\
 & \le\stackrel[{\scriptstyle i=1}]{{\scriptstyle n}}{\sum}I\left(M_{0}M'_{0};Y_{2,i}\vert Y_{2,i+1}^{\,\,n}\right)+H\left(\phi\left(Y_{1}^{n}\right)\right)\nonumber \\
 & \le\stackrel[{\scriptstyle i=1}]{{\scriptstyle n}}{\sum}I\left(M_{0}M'_{0}X_{i}Y_{2,i+1}^{\,\,n};Y_{2,i}\right)+nC_{1}\nonumber \\
 & =\stackrel[{\scriptstyle i=1}]{{\scriptstyle n}}{\sum}I\left(X_{i};Y_{2,i}\right)+nC_{1}\\
\nonumber \\
 & n\left(R_{0}+R'_{0}-\epsilon{}_{1,n}\right)\le I\left(M_{0}M'_{0};Y_{1}^{n}\right)\nonumber \\
 & =\stackrel[{\scriptstyle i=1}]{{\scriptstyle n}}{\sum}I\left(M_{0}M'_{0};Y_{1,i}\vert Y_{1}^{i-1}\right)\nonumber \\
 & \le\stackrel[{\scriptstyle i=1}]{{\scriptstyle n}}{\sum}I\left(X_{i}M_{0}M'_{0}Y_{1}^{i-1};Y_{1,i}\right)\nonumber \\
 & =\stackrel[{\scriptstyle i=1}]{{\scriptstyle n}}{\sum}I\left(X_{i};Y_{1,i}\right)
\end{flalign}
\begin{align}
 & n\left(R_{0}+R'_{0}-\epsilon'{}_{2,n}-\epsilon{}_{1,n}\right)\nonumber \\
 & \le I\left(M_{0};Y_{2}^{n}\right)+I\left(M'_{0};Y_{1}^{n}\right)\nonumber \\
 & \le I\left(M_{0};Y_{2}^{n}\right)+I\left(M'_{0};Y_{1}^{n}\vert M_{0}\right)\nonumber \\
 & =\stackrel[{\scriptstyle i=1}]{{\scriptstyle n}}{\sum}I\left(M_{0};Y_{2,i}\vert Y_{2,i+1}^{\,\,n}\right)+\stackrel[{\scriptstyle i=1}]{{\scriptstyle n}}{\sum}I\left(M'_{0};Y_{1,i}\vert M_{0}Y_{1}^{i-1}\right)\nonumber \\
 & \le\stackrel[{\scriptstyle i=1}]{{\scriptstyle n}}{\sum}I\left(M{}_{0}Y_{2,i+1}^{\,\,n};Y_{2,i}\right)+\nonumber \\
 & +\stackrel[{\scriptstyle i=1}]{{\scriptstyle n}}{\sum}I\left(M'_{0}Y_{2,i+1}^{\,\,n};Y_{1,i}\vert M_{0}Y_{1}^{i-1}\right)\nonumber \\
 & =\stackrel[{\scriptstyle i=1}]{{\scriptstyle n}}{\sum}I\left(M{}_{0}Y_{2,i+1}^{\,\,n};Y_{2,i}\right)+\stackrel[{\scriptstyle i=1}]{{\scriptstyle n}}{\sum}I\left(Y_{2,i+1}^{\,\,n};Y_{1,i}\vert M_{0}Y_{1}^{i-1}\right)+\nonumber \\
 & +\stackrel[{\scriptstyle i=1}]{{\scriptstyle n}}{\sum}I\left(M'_{0};Y_{1,i}\vert M_{0}Y_{1}^{i-1}Y_{2,i+1}^{\,\,n}\right)\nonumber \\
 & \overset{\left(a\right)}{=}\stackrel[{\scriptstyle i=1}]{{\scriptstyle n}}{\sum}I\left(M{}_{0}Y_{2,i+1}^{\,\,n};Y_{2,i}\right)+\stackrel[{\scriptstyle i=1}]{{\scriptstyle n}}{\sum}I\left(Y_{1}^{i-1};Y_{2,i}\vert M_{0}Y_{2,i+1}^{\,\,n}\right)+\nonumber \\
 & +\stackrel[{\scriptstyle i=1}]{{\scriptstyle n}}{\sum}I\left(X_{i}M'_{0};Y_{1,i}\vert M_{0}Y_{1}^{i-1}Y_{2,i+1}^{\,\,n}\right)\nonumber \\
 & =\stackrel[{\scriptstyle i=1}]{{\scriptstyle n}}{\sum}I\left(M{}_{0}Y_{2,i+1}^{\,\,n}Y_{1}^{i-1};Y_{2,i}\right)+\nonumber \\
 & +\stackrel[{\scriptstyle i=1}]{{\scriptstyle n}}{\sum}I\left(X_{i};Y_{1,i}\vert M_{0}Y_{1}^{i-1}Y_{2,i+1}^{\,\,n}\right)
\end{align}
\end{subequations} where $\left(a\right)$ is due to Csiszar sum
identity and the fact that $X_{i}$ is a deterministic function of
the messages. Finally define $U_{i}=\left(M_{0}Y_{2,i+1}^{\,\,n}Y_{1}^{i-1}\right)$
in  \eqref{eq:UBcommonProof} and then a time-sharing random variable
uniformly distributed $Q\sim Uniform\left[1:n\right]$ independent
of all R.V, and taking the limit $n\rightarrow\infty$ to complete
the converse proof.

\subsection{Proof for the AWGN BC}

The proof follows by the same arguments of the capacity proof for
the AWGN BC. Observe that this channel is stochastically degraded,
i.e. there exist $Y'_{1},Y'_{2}$ such that $P_{Y_{1}Y_{2}\vert X}=P_{Y'_{1}Y'_{2}\vert X}$
where
\begin{eqnarray*}
Y_{1}^{'}=X+Z_{1}^{'} &  & Z_{1}^{'}\thicksim{\cal N}\left(0,N_{1}\right)\\
Y_{2}^{'}=Y_{1}^{'}+\tilde{Z}{}_{2} &  & \tilde{Z}{}_{2}\thicksim{\cal N}\left(0,N_{2}-N_{1}\right)
\end{eqnarray*}
There are five inequalities in \eqref{eq:IB}, but as the channel
is stochastically degraded we have $I\left(U;Y_{2}\right)\le I\left(U;Y_{1}\right)$
and (\ref{eq:IB4}) which makes inequality (\ref{eq:IB5}) redundant.
Denote by $h\left(\cdot\right)$ the differential entropy of a continuous
R.V, and define the Gaussian entropy function and capacity by:
\begin{eqnarray*}
{\cal H}\left(x\right)={\textstyle \frac{1}{2}}\log\left(2\pi ex\right) &  & {\cal C}\left(x\right)={\textstyle \frac{1}{2}}\log\left(1+x\right)
\end{eqnarray*}
We bound $h\left(Y_{2}\vert U\right)$:
\begin{gather*}
{\cal H}\left(N_{2}\right)=h\left(Z_{2}\right)\le h\left(Y_{2}\vert U\right)\le h\left(Y_{2}\right)\le{\cal H}\left(P+N_{2}\right)
\end{gather*}
so there exist $\bar{\alpha}_{0}\in\left[0,1\right]$ such that
\begin{equation}
h\left(Y_{2}\vert U\right)={\cal H}\left(\bar{\alpha}_{0}P+N_{2}\right)\label{eq:Gauss_h(Y2|U)}
\end{equation}
where we use $\bar{\alpha}_{0}=1-\alpha_{0}$ for convenience of notation.
We can now bound $R_{0}$:
\begin{eqnarray*}
R_{0} & \le & I\left(U;Y_{2}\right)=h\left(Y_{2}\right)-h\left(Y_{2}\vert U\right)\\
 & \le & {\cal H}\left(P+N_{2}\right)-{\cal H}\left(\bar{\alpha}_{0}P+N_{2}\right)={\cal C}\left({\textstyle \frac{\alpha_{0}P}{N_{2}+\bar{\alpha}_{0}P}}\right)
\end{eqnarray*}
In order to prove inequality (\ref{eq:GCapacity4}), we use the scalar
EPI: 
\begin{align}
h\left(Y_{2}\vert U\right) & =h\left(Y_{2}^{'}\vert U\right)=h\left(Y_{1}^{'}+\tilde{Z}_{2}\vert U\right)\nonumber \\
 & \ge\frac{1}{2}\log\left(2^{2h\left(Y_{1}^{'}\vert U\right)}+2^{2h\left(\tilde{Z}_{2}\vert U\right)}\right)\nonumber \\
 & =\frac{1}{2}\log\left(2^{2h\left(Y_{1}^{'}\vert U\right)}+2^{2h\left(\tilde{Z}_{2}\right)}\right)\nonumber \\
 & =\frac{1}{2}\log\left(2^{2h\left(Y_{1}^{'}\vert U\right)}+2\pi e\left[N_{2}-N_{1}\right]\right)\label{eq:GaussEPI_h(Y2|U)}
\end{align}
From (\ref{eq:Gauss_h(Y2|U)}) and (\ref{eq:GaussEPI_h(Y2|U)}) we
have $h\left(Y_{1}^{'}\vert U\right)\le{\cal H}\left(\bar{\alpha}_{0}P+N_{1}\right)$.
Thus we can bound $R'_{0}+R_{1}$: 
\begin{eqnarray*}
R'_{0}+R_{1} & \le & I\left(X;Y_{1}|U\right)=h\left(Y_{1}\vert U\right)-h\left(Y_{1}\vert X\right)\\
 & = & h\left(Y_{1}^{'}\vert U\right)-h\left(Z_{1}\right)\\
 & \le & {\cal H}\left(\bar{\alpha}_{0}P+N_{1}\right)-{\cal H}\left(N_{1}\right)={\cal C}\left({\textstyle \frac{\bar{\alpha}_{0}P}{N_{1}}}\right)
\end{eqnarray*}
Similarly we bound $h\left(Y_{2}\vert UV\right)$:
\begin{gather*}
{\cal H}\left(N_{2}\right)=h\left(Z_{2}\right)\le h\left(Y_{2}\vert UV\right)\le h\left(Y_{2}|U\right)={\cal H}\left(\bar{\alpha}_{0}P+N_{2}\right)
\end{gather*}
so there exist $\alpha_{1}\in\left[0,\bar{\alpha}_{0}\right]$ such
that
\begin{equation}
h\left(Y_{2}\vert UV\right)={\cal H}\left(\alpha_{1}P+N_{2}\right)\label{eq:Gauss_h(Y2|UV)}
\end{equation}
We proceed by bounding the $R'_{0}$:
\begin{eqnarray*}
R'_{0} & \le & I\left(V;Y_{2}|U\right)+C_{1}=h\left(Y_{2}\vert U\right)-h\left(Y_{2}\vert UV\right)+C_{1}\\
 & = & {\cal H}\left(\bar{\alpha}_{0}P+N_{2}\right)-{\cal H}\left(\alpha_{1}P+N_{2}\right)+C_{1}\\
 & = & {\cal C}\left({\textstyle \frac{\left[\bar{\alpha}_{0}-\alpha_{1}\right]P}{N_{2}+\alpha_{1}P}}\right)+C_{1}\\
 & \triangleq & {\cal C}\left({\textstyle \frac{\alpha_{0}^{'}P}{N_{2}+\alpha_{1}P}}\right)+C_{1}
\end{eqnarray*}
where we define $\alpha_{0}^{'}\triangleq\left[\bar{\alpha}_{0}-\alpha_{1}\right]$,
thus $\alpha_{0}+\alpha_{0}^{'}+\alpha_{1}=1$. To prove inequality
(\ref{eq:GCapacity3}), we use again the scalar EPI:
\begin{align}
h\left(Y_{2}\vert UV\right) & =h\left(Y_{2}^{'}\vert UV\right)=h\left(Y_{1}^{'}+\tilde{Z}_{2}\vert UV\right)\nonumber \\
 & \ge\frac{1}{2}\log\left(2^{2h\left(Y_{1}^{'}\vert UV\right)}+2^{2h\left(\tilde{Z}_{2}\vert UV\right)}\right)\nonumber \\
 & =\frac{1}{2}\log\left(2^{2h\left(Y_{1}^{'}\vert UV\right)}+2^{2h\left(\tilde{Z}_{2}\right)}\right)\nonumber \\
 & =\frac{1}{2}\log\left(2^{2h\left(Y_{1}^{'}\vert UV\right)}+2\pi e\left[N_{2}-N_{1}\right]\right)\label{eq:GaussEPI_h(Y2|UV)}
\end{align}
From (\ref{eq:Gauss_h(Y2|UV)}) and (\ref{eq:GaussEPI_h(Y2|UV)})
we have $h\left(Y_{1}^{'}\vert UV\right)\le{\cal H}\left(\alpha_{1}P+N_{1}\right)$.
Thus we can bound $R_{1}$: 
\begin{eqnarray*}
R_{1} & \le & I\left(X;Y_{1}|UV\right)=h\left(Y_{1}\vert UV\right)-h\left(Y_{1}\vert X\right)\\
 & = & h\left(Y_{1}^{'}\vert UV\right)-h\left(Z_{1}\right)\\
 & \le & {\cal H}\left(\alpha_{1}P+N_{1}\right)-{\cal H}\left(N_{1}\right)={\cal C}\left({\textstyle \frac{\alpha_{1}P}{N_{1}}}\right)
\end{eqnarray*}
For the achievability we define three independent zero-mean Gaussian
random variables $U,V,W$ with variances $\alpha_{0}P,\alpha_{0}^{'}P,\alpha_{1}P$
respectively. Define $X=U+V+W$ and plug in inequalities \eqref{eq:IB}
to get inequalities \eqref{eq:GCapacity}.

\emph{Remark. }The capacity region is represented as a union of regions
defined by three parameters $\alpha_{0},\alpha_{0}^{'},\alpha_{1}$,
although each region is determined by only two of them as $\alpha_{0}+\alpha_{0}^{'}+\alpha_{1}=1$.
The reason becomes clear if one observes that each of those $\alpha$'s,
represents the power $\alpha P$ dedicated to each message. For example,
inequality (\ref{eq:GCapacity1}) stands for the decoding procedure
of decoder 2, when $m_{0}$ is the first (and sometimes the only)
message to decode. Thus, treating the power of the messages $m'_{0},m_{1}$
as additional noise, set the SNR to be ${\textstyle \frac{\alpha_{0}P}{N_{2}+\left[\alpha_{0}^{'}+\alpha_{1}\right]P}}$.

\bibliographystyle{IEEEtran}
\bibliography{BCUC}

\end{document}